\documentclass[a4paper]{article}

\usepackage[english]{babel}
\usepackage[utf8x]{inputenc}
\usepackage[T1]{fontenc}
\usepackage{breakcites}
\usepackage{caption}
\usepackage{color}
\usepackage{multirow}
\usepackage{makecell}
\usepackage{float}

\usepackage[a4paper,top=3cm,bottom=2cm,left=3cm,right=3cm,marginparwidth=1.75cm]{geometry}

\usepackage{amsmath}
\usepackage{graphicx}
\usepackage[colorinlistoftodos]{todonotes}
\usepackage[colorlinks=true, allcolors=blue]{hyperref}

\title{Imaging with quantum states of light}
\author{P.-A.~Moreau$^*$, E.~Toninelli, T.~Gregory and M.~J.~Padgett$^*$\\SUPA, School of Physics and Astronomy, University of Glasgow, Glasgow, G12 8QQ, UK\\~\\$^*$  e-mail: paul-antoine.moreau@glasgow.ac.uk; miles.padgett@glasgow.ac.uk}

\begin{document}
\maketitle

\begin{abstract}The production of pairs of entangled photons simply by focusing a laser beam onto a crystal with a non-linear optical response was used to test quantum mechanics and to open new approaches in imaging. The development of the latter was enabled by the emergence of single photon sensitive cameras able to characterize spatial correlations and high-dimensional entanglement. Thereby new techniques emerged such as the ghost imaging of objects --- where the quantum correlations between photons reveal the image from photons that have never interacted with the object --- or the imaging with undetected photons by using non-linear interferometers. Additionally, quantum approaches in imaging can also lead to an improvement in the performance of conventional imaging systems. These improvements can be obtained by means  of image contrast, resolution enhancement that exceed the classical limit and  acquisition of sub-shot noise phase or amplitude images. In this review we discuss the application of quantum states of light for advanced imaging techniques. 
\end{abstract}

\section*{Introduction}
\begin{figure}[H]
\centering
\includegraphics[width=0.8\textwidth]{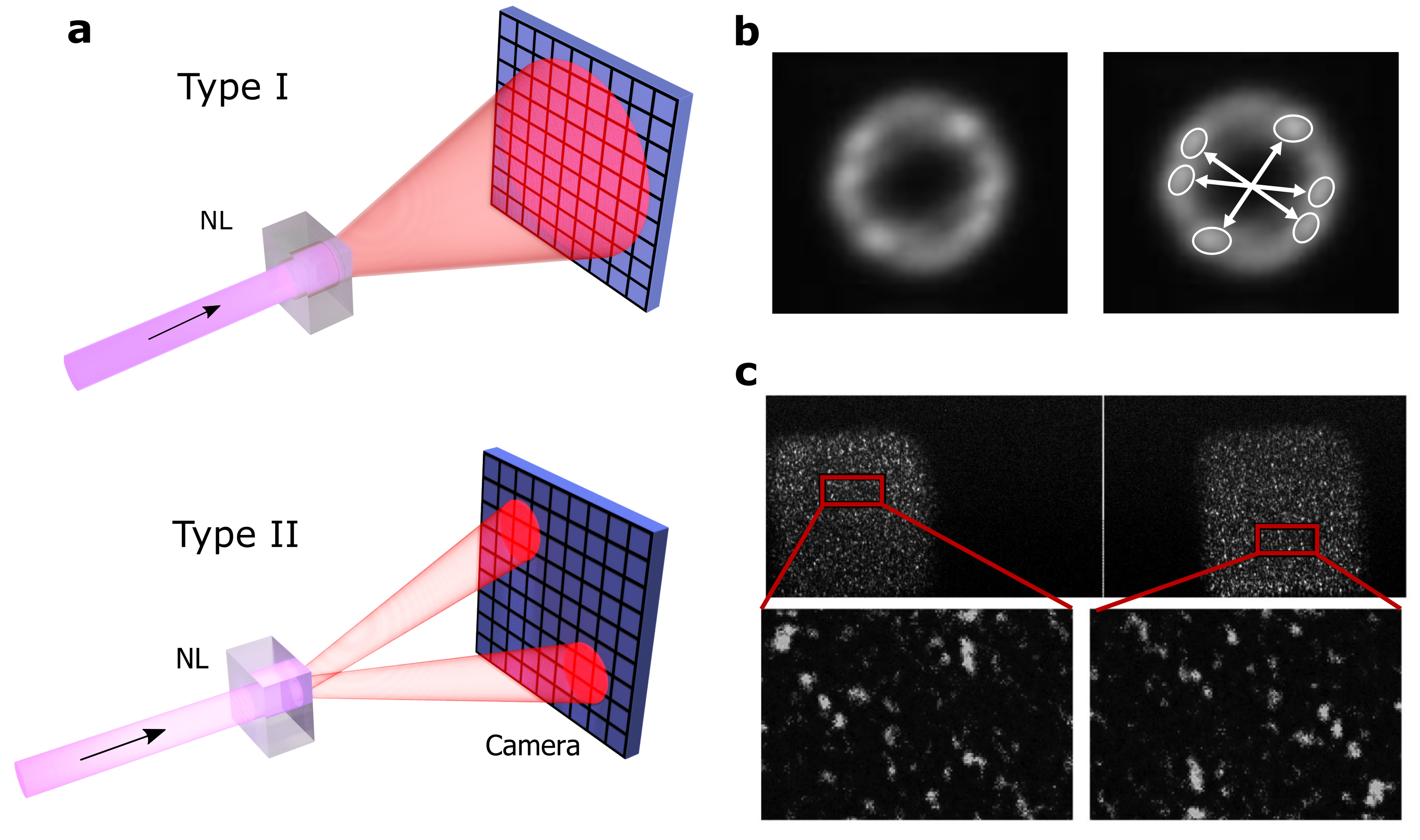}
\caption{\textbf{Generation and detection of quantum correlations in spontaneous parametric down-conversion (SPDC).} \textbf{a} SPDC generation in a non-linear crystal (NL). The process is depicted here for Type I phase matching (top) and Type II phase matching (bottom), leading to the emission of one and two SPDC beams respectively. Inside these beams spatial photon correlations can be detected. \textbf{b} Observation of correlations within a type I SPDC beam in a high gain regime. The image on the left is reproduced on the right with white arrows highlighting the correspondence between similar intensity features that are due to intensity correlations within diametrically opposed portions of the beam. \textbf{c} Observation of correlations between type II SPDC beams in a high gain regime. The left and right images correspond respectively to the two beams (signal and idler) imaged on different regions of a camera. One can also observe similar patterns within the two beams with a 180$^\circ$ rotation, which is a signature of momentum anti-correlations. Panel b is reproduced from REF.~\cite{devaux2000spatial}, Springer Nature Limited. Panel c is reproduced from REF.~\cite{jedrkiewicz2004detection}, APS.}
\label{fig:downconversion}
\end{figure}

With the emergence of modern non-linear optics in the second half of the 20th century~\cite{franken1961generation,wang1965measurement,giordmaine1965tunable}, physicists found the source of choice to conduct the desired tests of quantum mechanics. Light, unlike other physical systems, remains well isolated from its environment and is, therefore, by its nature not very sensitive to the effects of quantum decoherence. This good insulation of photons from their environment and from each other is highly desirable in order to study or harness the quantum properties of a system; however, it is also a drawback when it comes to the production of entangled particles because it is difficult to make two photons interact with each other to create entanglement. In early tests of quantum mechanics principles, the photons were generated through a cascaded two photon emission process from single atoms \cite{kocher1967polarization,freedman1972experimental,aspect1982experimental}. Such techniques are rather difficult to implement. They were quickly superseded by the use of non-linear optics that allows the creation of twin photons within a medium with a non-linear response such as a non-linear crystal. It was shown theoretically~\cite{louisell1961quantum,haus1962quantum,klyshko1969scattering} and experimentally~\cite{harris1967observation,akhmanov1967quantum,burnham1970observation} that it is possible to generate photon pairs through the interaction of a single pump photon with a non-linear medium. Such a three wave interaction process between a pump photon and two lower frequency -- signal and idler -- photons is called spontaneous parametric down-conversion (SPDC). It was shown that SPDC allows the generation of quantum states of light~\cite{ghosh1987observation,ou1988violation}. In particular, it is possible to generate entanglement in polarisation~\cite{kwiat1995new} using such a non-linear process ~\cite{shih2003entangled,genovese2005research}. SPDC has since been widely used as a source for a variety of fundamental demonstrations of quantum mechanics and quantum information protocols. Notably, the parametric down conversion process has been used in the demonstration of Hong-Ou-Mandel (HOM) two photon interference~\cite{hong1987measurement}, implementation of delayed-choice experiments~\cite{kim2000delayed,ma2016delayed}, quantum teleportation~\cite{ren2017ground}, elaboration of optical quantum gates and information protocols~\cite{o2003demonstration,qiang2018large} as well as in the entanglement based on quantum key distribution~\cite{lo2014secure}. It was also used in two realisations of the loophole free Bell test experiment~\cite{PhysRevLett.115.250402,PhysRevLett.115.250401}.\\
Interestingly, as demonstrated for imaging applications, the light emitted through the SPDC process (Fig. \ref{fig:downconversion}) exhibits quantum correlations in both position and momentum~\cite{howell2004realization}. The existence of momentum correlations between the photons created and annihilated in the SPDC process, was highlighted experimentally as early as 1970 (Ref. ~\cite{burnham1970observation} together with the first demonstration of the existence of temporal photon correlations in the light emitted in this process. These correlations were perfectly expected as they are a result of momentum conservation between the annihilated pump photon and the two photons emitted in the SPDC process. Even more interestingly, it was recognized that the state produced through the SPDC process is in fact a good approximation of the original Einstein-Podolsky-Rosen (EPR) state of entanglement~\cite{einstein1935can} because it presents both position and momentum correlations~\cite{klyshko1988photons}. The availability of such a state and the simplicity of its generation -- requiring only to pump a non-linear crystal with a laser -- initiated the development of new types of imaging experiments. Through these experiments emerged the field of 'quantum imaging'. In the following we will give an overview of how the quantum behaviour of light can be detected through imaging and how it can be harnessed advantageously in imaging protocols. In particular, we will describe why SPDC sources play an essential role in these realizations. 

\section*{Cameras to detect quantum behaviour}
The widespread use of SPDC as a source of quantum states rapidly generated interest in the study of correlations and entanglement in the spatial domain. Not only did it promise to develop new types of imaging, but it also allowed to engineer very high dimensional quantum states. The concurrent development of new types of camera technologies enabled the detection of single photons with cameras~\cite{jost1998spatial,basden2003photon}. The new types of quantum optical demonstrations no longer relied on scanning point-like avalanche photo-diodes but rather on spatially resolved detectors. This development in detection techniques enabled highly parallel correlation measurements, and ultimately led to time efficient detection of quantum signatures that may be exploited in the context of quantum information protocols. In this section we describe different camera technologies that have led to quantum imaging demonstrations of fundamental nature; we highlight different regimes in which these cameras can perform and discuss the respective advantages and disadvantages of the current technology.

\subsection*{Detection of quantum correlations}

Following the early detection of spatial photon correlations~\cite{burnham1970observation}, further work was performed throughout the 1990's both to characterize and exploit the quantum correlations emitted through SPDC ~\cite{rarity1990experimental,pittman1995optical,strekalov_observation_1995}. However, these techniques were inherently inefficient because they used avalanche photo-diodes to detect single photons, and a scanning pinhole to detect the spatial features of the correlations. Such a pinhole filtering technique leads to the loss of the vast majority of the photons, and therefore, the experiment requires a long time to measure the spatial form of the correlation. Since that time researchers have tried to detect spatial correlations between photon pairs with cameras, thereby removing the necessity to filter at one particular position.
The first attempt of detecting spatial correlations of quantum origin with a camera was performed in 1998 (Ref.~\cite{jost1998spatial}) using a photon-counting intensified CCD camera (ICCD). The results of this work --- despite the relatively noisy images of photon detection --- inspired many experiments with more technologically advanced cameras.\\ 
However, because of the technological limitations including the dark count, the noise and low quantum efficiencies of available cameras the subsequent tests had to be performed under a different regime. In 2000 a conventional single-frame camera was used to detect intensity correlations in the spatial spectrum of SPDC beams~\cite{devaux2000spatial}. A lithium triborate crystal was used to generate a down-converted beam in a high-gain regime. To ensure an efficient emission of the signal and idler waves through parametric fluorescence, the idler and the signal phases must match throughout the propagation of the waves inside the non-linear material. This phase matching ensures that the waves emitted at some point of the crystal do not interfere destructively with the upstream emissions. There are two main types of phase matching that exists in the context of parametric fluorescence, type I and type II. In type I phase matching both signal and idler waves are emitted with the same polarization and therefore propagate along the same birefringent axis of the non-linear crystal~\cite{couteau2018spontaneous,boyd2003nonlinear}. In type II, however, the two waves that are emitted have orthogonal polarizations and propagate along two distinct birefringent axis which permit, under non-collinear phase matching conditions the generation of two distinct beams propagating in different directions. 
Under the conditions used in~\cite{devaux2000spatial}, the relatively bright beam intensity is able to exceed the noise floor of the camera. In this work, the correlated intensity fluctuations were detected within different parts of the beam ({Fig. \ref{fig:downconversion} b}).\\



In the aforementioned demonstration the correlations were simply observed and spatially characterized. However, it is also possible to demonstrate sub-shot noise behaviour with such correlations, thus establishing their quantum nature. Within two correlated regions of interest (ROI) of the beam, the detected intensities are indeed expected to follow the same temporal fluctuations. If these joint fluctuations are due to the arrival of correlated photons rather than a result of classical intensity fluctuations, then, for an ideal system, one would expect to detect in both ROIs exactly the same number of emitted photons. As a consequence, by subtracting the two signal intensities one may obtain zero and the fluctuation of this intensity difference will also tend to zero. By contrast, classically correlated intensities when subtracted cannot go below a certain limit. This limit, called shot noise, is due to the quantum nature of light and the fact that the number of photons in a light beam is subject to a fundamental standard deviation which is equal to the square root of the average number according to Poisson statistics. The shot noise limit corresponds to the intensity fluctuations of the lowest noise classical state: a coherent state which is that of an ideal laser. If by subtracting two intensity signals one obtains a quantity that fluctuates less than that of the shot noise, then the underlying statistics is said to be sub-shot noise and one can conclude that the two beams exhibit quantum correlations~\cite{jakeman1985analysis,jakeman1986use}. This was achieved in parametric down-conversion in the context of correlated single mode beams~\cite{rarity1987observation,heidmann1987observation,nabors1990two,tapster1991sub,ribeiro1997sub,bondani2007sub,iskhakov2009generation} before being demonstrated in the context of imaging. Indeed, a few years after the aforementioned demonstration~\cite{devaux2000spatial}, such a quantum signature of correlations in images was observed under a similar regime~\cite{jedrkiewicz2004detection,jedrkiewicz2006quantum}. It was done using parametric fluorescence generated in a barium borate (BBO) crystal with type II phase matching. By acquiring images of the bright fluorescence beams, the authors showed the sub-shot noise nature of the detected spatial correlations (see Fig. \ref{fig:downconversion} c).\\
As mentioned above, although it is true that these results demonstrated genuine photon quantum correlations, they were performed in a high-gain regime of the down-converted emission. This means that such correlations were not composed of pure twin photon correlations but also of higher order photon correlations generated through the stimulated emission. Such higher order terms can introduce excess noise when the imaging system is subject to losses that is in non-ideal conditions. A few years later the sub-shot noise behaviour of SPDC light in images captured with an electron-multiplying CCD camera (EMCCD) was demonstrated~\cite{blanchet2008measurement}.
Moreover, in another work~\cite{brambilla2008high}, a scheme was proposed to use photon correlations to achieve sub-shot noise imaging, that is the acquisition of images in a scheme that outperforms classical imaging schemes in terms of noise. It was suggested that an optimal regime to use for such a realization is a bright light low gain regime, which prevents the introduction of excess noise. Following these suggestions, a similar regime was later used to detect twin photons~\cite{brida2009measurement} on the way to realizing sub-shot noise imaging of a low transmission sample~\cite{brida2010experimental}. To access this particular regime, a pump laser with pulse duration much longer than the coherence time of the SPDC was used to ensure that the number of photons emitted per spatio-temporal mode of fluorescence emission was low enough to make the stimulated emission negligible. Hence a very low contribution from the higher order photon number correlations was ensured, thus limiting the impact of the excess noise. This low emission rate led to a demonstration of sub-shot-noise spatial correlations without background subtraction using a conventional CCD camera~\cite{brida2010experimental}.

\subsection*{Efficient characterization with cameras}
The aforementioned demonstrations were focused on simply detecting a signature of quantum correlations. As we have seen, this can be done in a bright light regime with conventional scientific cameras. However, such cameras are not sensitive enough to detect single photons. Indeed the noise floor of such cameras is typically of several electrons even when the sensor is cooled. Under such conditions the detection of a single photon leading to a photo-electron trapped in the CCD well would not be observable in the image, which would be largely dominated by the technical noise of the camera. This considerably limits the range of quantum behaviour that can be observed with such cameras. To detect more subtle quantum characteristics that require the detection of single photons one needs to use different camera, for example, EMCCD. In contrast to conventional CCDs, EMCCD cameras incorporate on-chip gain placed before the charge reading stage~\cite{jerram2001llccd}. The gain register generates a multiplicative avalanche effect that occurs in around 500 stages. At each stage the electrons contained in accelerated potential wells have a small probability to generate a secondary electron through impact ionization with the chip substrate. This amplification before reading has the potential to make single photon events to emerge from the camera readout noise.\\
With such cameras it is possible to develop photon-counting strategies~\cite{basden2003photon}. As mentioned before, these strategies led to the detection of sub-shot noise features in SPDC light~\cite{lantz2008multi,blanchet2008measurement}. It was followed by several other demonstrations aimed at detecting and using quantum correlations with EMCCD cameras ~\cite{zhang2009characterization,blanchet2010purely,toninelli_sub-shot-noise_2017,reichert2017biphoton,reichert2018massively}.

\begin{figure}[H]
\centering
\includegraphics[width=\textwidth]{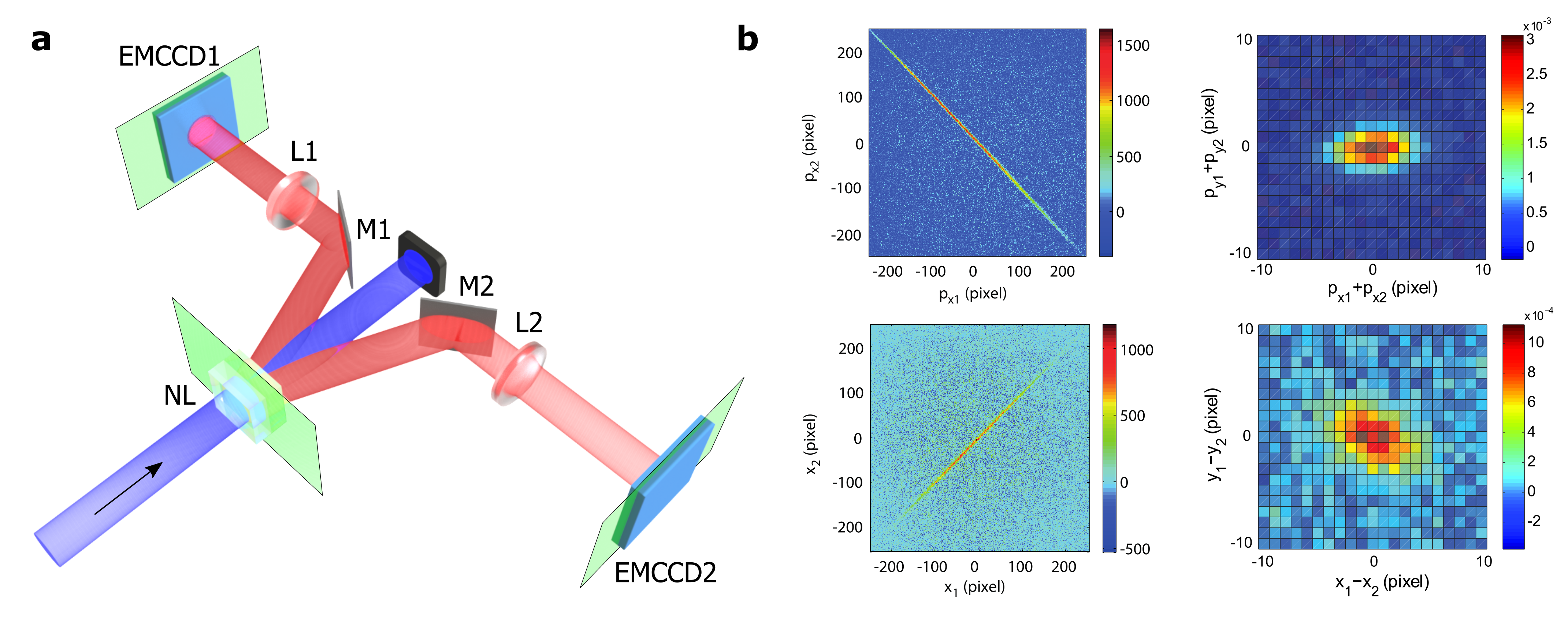}
\caption{\textbf{Experimental test of Einstein-Podolsky-Rosen (EPR) paradox in images.} \textbf{a} Experimental setup used to demonstrate a two dimensional EPR paradox with a pair of synchronised electron multiplying CCD cameras (EMCCD1 and EMCCD2). A nonlinear crystal (NL) cut for type II downconversion is pumped with a UV laser, the two spatially separated beams are then sent to two cameras, either by reimaging the crystal plane as shown here (the green planes indicate the crystal plane and the crystal image planes) or by imaging a Fourier plane of the crystal to detect the momenta of the photons. $M1$ and $M2$ are mirrors, $L1$ and $L2$ are lenses and BD is a beam dump. The pump is represented by a blue beam and the signal and idler are represented by red beams. $x_i$ and $y_i$ are the transverse positions of detection of photon $i=\{1,2\}$ along the horizontal ($x$) and vertical ($y$) directions. $p_{xi}$ and $p_{yi}$ are the transverse momenta of photon $i=\{1,2\}$ along the horizontal and vertical directions, respectively. \textbf{b} Detected conditional probability distribution of bi-photons in the acquired images. The results exhibit quantum correlations in both momentum (top) and position (bottom). The right column corresponds to the cross-correlations of the images composed of single photon events extracted from the two cameras. The correlation peaks are the signature of two dimensional quantum correlations that can be shown to demonstrate a high dimensional EPR paradox. Panel b is reproduced from REF.~\cite{moreau2014einstein}, APS.}
\label{fig:EPR}
\end{figure}

With single-photon cameras one can also demonstrate more fundamental quantum phenomena such as an ERP paradox~\cite{einstein1935can}. Correlations demonstrating the EPR paradox were performed in a series of experiments~\cite{devaux2012towards,moreau2012realization,edgar2012imaging,moreau2014einstein} ({Fig. \ref{fig:EPR} }).

In these experiments SPDC light is used to produce EPR-type spatial correlations. The correlated beams are then detected with an EMCCD camera and correlations are recorded between different regions of the camera. These correlations are measured in two different configurations: the first one in which the camera acquires images of the crystal plane thus recording correlations between the positions of the down-converted photons and the second one where a Fourier plane of the crystal is imaged onto the camera thus leading to the detection of correlations between the transverse momenta of the photons. The existence of such correlations in both position and momentum is demonstrative of the EPR paradox. In contrast to the previous manifestation of the EPR paradox~\cite{howell2004realization}, cameras enable an inherently more efficient demonstration: scanning a point-like single photon detector is inefficient as it leads to the loss of most of the photons.\\
Two synchronized cameras were also used in~\cite{moreau2014einstein} to ensure the full spatial separation of the photons. In this experiment the sub-shot noise nature of the correlations was simultaneously detected. Cameras allow the characterisation of very high dimensional quantum correlation features~\cite{edgar2012imaging,moreau2014einstein} through the great number of pixels they posses (typically $\sim 300k-1M$ pixels) but also they allow the opportunity to perform highly parallel detections of such quantum correlations. Using a similar scheme, the EPR paradox was shown with only a single pair of frames in each of the two configurations (position or momentum detection)~\cite{lantz2015einstein}. This signifies that no temporal averaging had to be performed in order to detect these features. Instead, by using single frames only, the number of synchronously recorded samples of the state in one frame was sufficient to perform the statistical averaging. The total acquisition time for such an acquisition requires only two exposures of $0.03s$ each.\\
Other demonstrations led to fast characterisation of correlations in the space domain using single pixel cameras~\cite{takhar2006new,duarte2008single}. This was done with a combination of a digital micromirror device and a single pixel single photon avalanche diode (SPAD)~\cite{howland2013efficient,schneeloch2013violation}. Using compressive sensing techniques relying on sparsity in spatial correlations between entangled photons, they were able to demonstrate a thousand fold improvement in the acquisition times over conventional raster scanning techniques (reduced from 310 days to 8 hours).

\subsection*{Intensified cameras}
Another type of camera that is useful in the context of quantum imaging is the intensified camera. Intensified cameras are usually scientific CCD or CMOS cameras that have an image intensifier placed before the sensor. The image intensifiers~\cite{lampton1981microchannel} are composed of a photo cathode that converts the photons into electrons, followed by a micro channel plate that multiplies the electrons through the influence of high voltage but very short --- of the order a few nanoseconds --- electrical pulses, and finally a phosphor screen that reconverts electrons back into photons. The photons emitted by the phosphor screen are then detected by the camera. With such an amplification happening before the detection, a light beam arriving into the intensifier --- that would be composed of only a single photon --- is converted into an amplified signal of many photons with very little additional noise due to the short time gates that limits not only the influence of the dark-current noise but also of incoming spurious light. One can then acquire images in which the huge spikes that appear correspond to single-photon detection events. By applying a detection threshold to such images one may obtain images of true single-photon detection events.\\

As mentioned before, the earliest use of an ICCD to record correlations of quantum origin was performed in 1998 (Ref.~\cite{jost1998spatial}). This demonstration inspired further measurements of the correlations ~\cite{abouraddy2001demonstration,pires2009direct}, until the technology became mature enough to lead to substantial use of such cameras in quantum imaging. Several new realisations used ICCDs to perform quantum imaging experiments.  EPR based imaging was demonstrated~\cite{aspden_epr-based_2013} by triggering an ICCD camera with a SPAD whose detection of a photon heralds the arrival of its entangled twin. The same year, real-time imaging of two photon entanglement in two modes was demonstrated using a similar acquisition setup~\cite{fickler2013real}. Another important quantum phenomenon was later evidenced through the use of an intensified camera when a shot by shot observation of a HOM effect was performed~\cite{jachura2015shot}. The authors had previously shown that the intensified CMOS camera they used was able to perform spatially resolved multiphoton counting~\cite{chrapkiewicz2014high}. This work led to subsequent demonstrations using a similar setup, such as the acquisition of holograms in a single photon regime~\cite{chrapkiewicz2016hologram}, the demonstration of bi-photon mode engineering for quantum-enhanced interferometry~\cite{jachura2016mode}, and the demonstration of a wave-vector multiplexed quantum memory for photons interacting with cold atoms~\cite{parniak2017wavevector}.

\subsection*{Other camera technologies}
We have seen in this section three main types of camera technologies that can be used to perform quantum imaging, in Table \ref{tab:cam} we have reported some of the most important characteristics of such cameras in the context of quantum imaging. It is useful to note that new technologies such as SPAD arrays~\cite{guerrieri2010two,guerrieri2010sub,veerappan2011160,gariepy2015single}, matrices of superconducting nanowires single photon detectors~\cite{miki201464,allman2015near}, or new back-side-illumination CMOS technologies~\cite{ma2017photon} are able to resolve the number of photons and are very promising alternatives to the cameras presented here. However, these technologies are not yet sufficiently mature to be implemented in quantum imaging applications involving correlated photons.

\begin{table}[H]
  \begin{center}
  \resizebox{\textwidth}{!}{
    \begin{tabular}{c|c|c|c}
& \textbf{Scientific cameras} & \textbf{EMCCD}& \textbf{Intensified Cameras}\\
\hline
\makecell{Overall detection\\ efficiency} & up to $\sim 95\%$ & \makecell{$\sim 60\%$ after thresholding;\\ Quantum efficiency $> 80\%$, (at $-90^oC$)} & $\sim 10-20\%$ after thresholding \\
      \hline
      \makecell{Lowest accessible\\ light regime 
      in photons\\ per pixels frames} & $> 1k-10k$ & $0.01-0.15$ & $10^{-3}-10^{-4}$\\
      \hline
      \makecell{Typical noise at  lowest\\ accessible light regime\\  per pixel par frame} & $1-5~e^-$ & $\sim 0.002$ equivalent photons after thresholding & \makecell{$\sim 10^{-4}$ equivalent photons after thresholding, \\that is, few photons per frame}\\
      \hline
Main utility in quantum imaging & Detecting intensity correlations & \makecell{Fast statistical\\ characterization of correlations\\ in photon-counting~\cite{lantz2014optimizing};\\ Relatively but relatively noisy\\ single photon pairs \\identification~\cite{tasca2013optimizing}} & \makecell{Low noise single\\ photons pairs identification} \\
    \end{tabular}}
  \end{center}
      \caption{Comparison of different camera technologies used in quantum imaging.}
      \label{tab:cam}
\end{table}

\section*{Improved imaging with quantum light}

So far we have seen how quantum imaging and especially the use of commercially available single-photon sensitive cameras can be used to efficiently detect correlations between photons in high spatial dimensions. These quantum proprieties can also be harnessed using similar techniques and tools with potentially improved performance compared to what can be accessible via purely classical schemes. We will here review the various works that have been conducted along these lines. For this review we will distinguish two different categories in image improvement targeted by researchers: the first is a reduction of the noise in the recorded phase or intensity of the images and the second is the improved resolution of such images.

\subsection*{Improved optical metrology.}

With the emergence of the concepts and the way to generate quantum squeezing~\cite{andersen201630,schnabel2017squeezed} and definite number of photons states~\cite{mandel1964physical,stoler1974photon,jakeman1986use}, new schemes were devised, able to harness quantum proprieties of light to reach levels of sensitivity beyond that achievable classically ~\cite{hollenhorst1979quantum,caves1981quantum,snyder1990sub,tapster1991sub}. In the context of optics a squeezed state can be defined as a state that presents a reduced uncertainty along one of the two field quadratures, compared to that of the quadratures for a coherent or vacuum state (state of an ideal laser). If the quadratures are bounded by the Heisenberg uncertainty principle, the uncertainty on each one is not fundamentally bounded. 

Another class of states present a reduced uncertainty compared to a coherent state. These states that have a definite number of photons are also called Fock states.
Compared to coherent states that  whose intensity is uncertain, the intensity of such states is perfectly defined. The Heisenberg trade-off being that the phase of a Fock state is perfectly uncertain. In both cases of squeezed and Fock states the reduced uncertainty they present can be exploited to increase the measurement sensitivity.\\
Such techniques can be included in the general domain of quantum metrology~\cite{giovannetti2004quantum,degen2017quantum}, that aim at using quantum properties to perform improved measurements. The domain was largely initiated by the desire to improve the sensitivity of gravitational wave detectors, which led to the early theoretical developments of quantum non-demolition measurement~\cite{caves1980measurement}, and the use of squeezed light in interferometric gravitational wave detectors~\cite{abadie2011gravitational,aasi2013enhanced}. 

\begin{figure}[H]
\centering
\includegraphics[width=\textwidth]{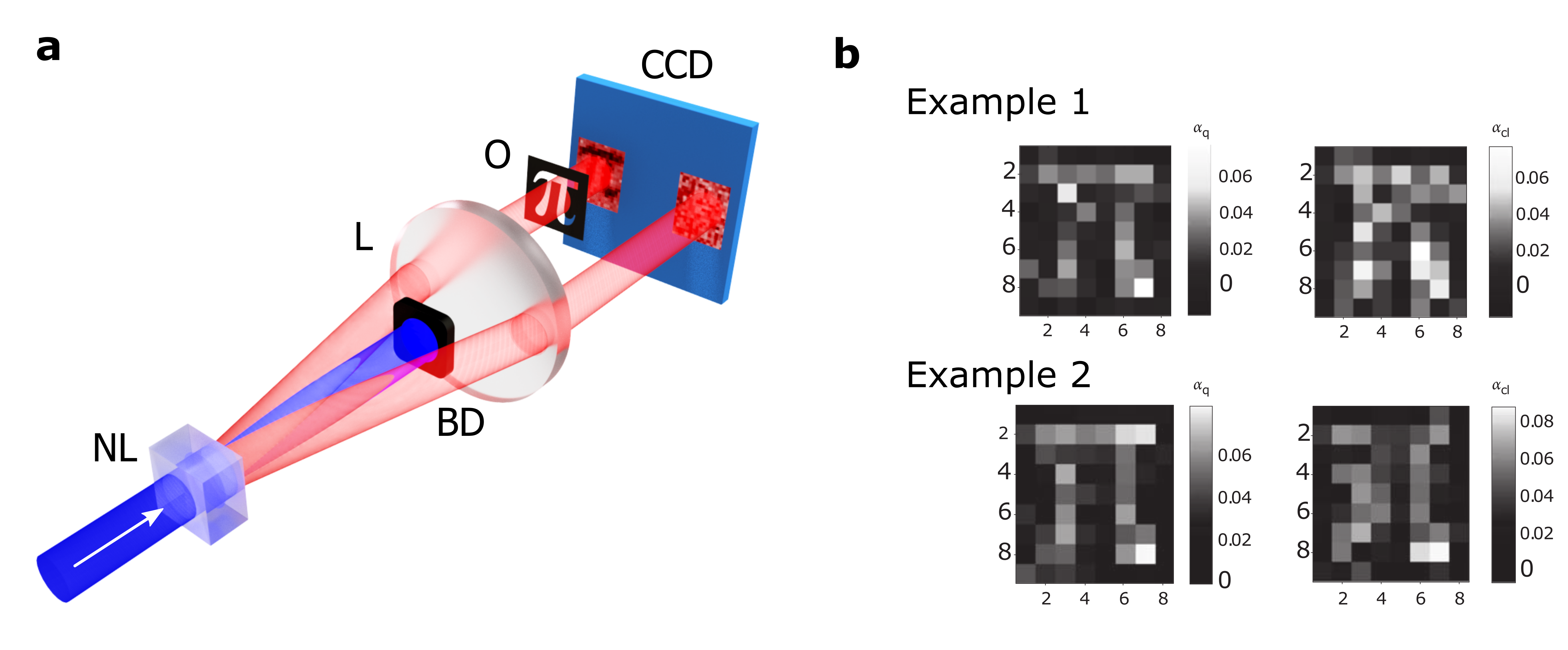}
\caption{\textbf{Sub-shot-noise quantum imaging.} \textbf{a} Quantum correlated beams are generated in a barium borate non-linear crystal (NL) through type II SPDC. One of the beams probes a low absorptive object (O) before being detected on a camera, the second beam is directly detected on a different part of the same camera. By subtracting the intensities detected within each of the beams it is possible to remove some of the shot noise in the acquired image of the object. BD is a beam dump. \textbf{b} Two examples of images acquired: the images on the left are obtained after subtraction, the ones on the right are the images recorded directly by the camera. One can see that the images on the left appear less noisy, which is due to the sub-shot-noise nature of the correlations between the beams. It was shown that such images themselves exhibit sub-shot-noise qualities. Panel b is reproduced from REF.~\cite{brida2010experimental}, Springer Nature Limited.}
\label{fig:brida}
\end{figure}

In the context of imaging, one application of quantum metrology arises in the measurement of delicate samples~\cite{taylor2016quantum,wolfgramm2013entanglement}. By enabling the use of sub-poissonian statistics of light and the possibility to surpass the classical noise limit, quantum metrology schemes allow to perform imaging of an object by exposing it to fewer photons than for a classical, shot-noise limited, measurement ~\cite{davidovich1996sub}.\\
For spatially single mode measurements, squeezing would appear as an opportunity to perform improved measurements over classical techniques in both phase~\cite{caves1980measurement,andersen201630} and absorption estimation~\cite{xiao1988detection}. The quantum advantage accessible through squeezing is indeed potentially significant as it is possible to produce states exhibiting squeezing as high as 15dB~\cite{vahlbruch2016detection}. The use of spatial squeezing in imaging could potentially lead to interesting applications~\cite{kolobov2000quantum,giovannetti2004quantum}. Thus far, spatially multimode squeezed states have only been used to improve beam localisation~\cite{treps2002surpassing,treps2003quantum} and in the detection of entanglement between a few spatial modes~\cite{boyer2008entangled,lassen2009continuous,wagner2008entangling}. The restricted application to date is mainly due to the difficulty in generating these states and building spatially resolved homodyne detectors.\\
An alternative form of quadrature squeezing suitable for absorption measurements is to use bright amplitude squeezed light states that can exhibit spatial correlations, such as the light emitted by an optical parametric oscillator (OPO) when running above threshold~\cite{heidmann1987observation}.
OPO allows to use the correlated nature of the signal and idler beams to improve the sensitivity of absorption measurements~\cite{snyder1990sub,ribeiro1997sub}. There are two limitations of such realizations. First, as in the case of phase quadrature squeezing, the technique is highly sensitive to the technical noise that is present at low frequencies in the source, and therefore, measurements need to be shifted to frequencies of a few MHz. The second limitation is due to the thermal nature of the down-converted beams, as the stimulated re-emission of the pairs is non-negligible when operating the OPO above-threshold that for a pump power is leading to a gain that is greater than the intra-cavity losses~\cite{heidmann1987observation}. In such a context, the metrology schemes become extremely sensitive to losses because a noisy thermal contribution is added to the estimated experimental parameter by reintroducing into the statistics thermal light that is no longer correlated. As a consequence, such schemes are usually limited to low absorption samples.\\
Another strategy to limit these drawbacks is to remain in a regime dominated by SPDC. Two techniques were proposed to use SPDC to perform sub-shot noise evaluation of the transmission of a sample for a single-mode measurement~\cite{jakeman1986use}. One of the two methods was based on the  feed forward method to generate Fock states to illuminate the sample. This method was recently implemented in single-mode transmission metrology~\cite{sabines2017sub}.  However, even more interestingly in the context of imaging, it was shown that one can use the same intensity correlated nature of SPDC beams to produce quantum improved estimation of a channel transmission with conventional --- non-photon-counting --- photo diodes~\cite{tapster1991sub}.\\
Such a regime of bright SPDC is similar to the regime proposed in~\cite{brambilla2008high} and exploited later~\cite{brida2009measurement} to detect and use quantum correlations with low noise conventional CCD cameras. Interestingly, the same regime was used again in the demonstration of the the first sub-shot noise imaging experiment~\cite{brida2010experimental} referred to above. In this remarkable realisation, type II SPDC beams pumped with long pulse duration were used to make the stimulated emission negligible. In this experiment, one of the two beams was sent through a low absorption object ($\sim 5\%$) imaged on the camera, while the second beam was propagated freely before being detected on a different part of the same camera in an optically equivalent plane as the first beam ({Fig. \ref{fig:brida}}).

Due to the presence of spatial correlations between the two beams, the shot-noise caused intensity fluctuations will be the same within the two beams at a particular transverse position $\bf{r}$. As a consequence, and because the object absorption was relatively low (preserving therefore the correlations between the beams), one can remove some of the measurement noise (instantaneous fluctuation) present in the first beam simply by subtracting the detected image intensity of the second beam, following the scheme proposed in~\cite{brambilla2008high}. By doing so, it is possible to obtain sub-shot noise images of a low absorption object~\cite{brida2010experimental}. The same approach was also implemented within a wide-field microscope~\cite{samantaray2017realization}. Yet, this subtraction method is limited to the imaging of low absorption objects because simply subtracting the intensity of the two beams to estimate the absorption of the object is a sub-optimal estimator that leads to a rapid loss of the quantum advantage for objects with higher absorption. However, another optimized estimator was reported~\cite{moreau2017demonstrating} that is less sensitive to losses. See~\cite{losero2018unbiased} for a systematic comparison of this new estimator to estimators previously used. This new estimator was used to demonstrate an absolute (unconditional) quantum advantage in absorption estimation for object presenting absorption up to $50\%$ in spatially single-mode measurements~\cite{moreau2017demonstrating}. Additionally, a heralded source of multi-mode mesoscopic sub-Poissonian light was shown through post selection of the detected intensities~\cite{iskhakov2016heralded} with the potential to bring previous transmission metrology demonstration to brighter regimes.\\
Using quantum states of light can also lead to improved estimation of the optical phase~\cite{giovannetti2004quantum,nagata2007beating}. The interferometric behaviour of NOON states can be used to increase the phase measurement sensitivity within an interferometer with Heisenberg scaling, that is an uncertainty on the phase estimation $\Delta\phi$ scaling as $\Delta\phi=1/N$ where N is the number of photons in the NOON state. This can be explained by the fact that the photonic de Broglie wavelength of such a state is equivalent to that of a single particle with wavelength $\lambda/N$ (Ref.~\cite{jacobson1995photonic,fonseca1999measurement}). The use of NOON states was implemented in the form of an interferometric microscope setup where a phase object was raster scanned to give an image with precision beating the standard quantum limit as dictated by the shot noise~\cite{ono2013entanglement}.\\
Another quantum interferometric technique that can reach Heisenberg-limited sensitivities, is the so called SU(1,1) non-linear interferometric method~\cite{yurke19862,leonhardt1994quantum,plick2010coherent}. In contrast to the NOON state interferometry, it presents an advantage that the quantum correlations are generated inside the interferometer, as the scheme is implemented as a Mach-Zehnder interferometer where the two beam-splitters are replaced by non-linear active elements pumped by the same source. This allows the preservation of a good quality of entanglement inside the interferometer and makes the method resilient to detection losses~\cite{marino2012effect,ou2012enhancement}. Experimental demonstrations of enhanced sensitivity have been performed~\cite{jing2011realization,hudelist2014quantum,manceau2017detection}. Such a scheme may also be implemented to perform Heisenberg limited phase imaging.

In addition to improving the fundamental limits of precision, quantum correlations can be harnessed to give technical ameliorations in imaging. For example, an implementation of a technique called 'heralded imaging' allows the use of quantum temporal correlations to trigger an intensified CCD camera for $\sim 4ns$ only when a signal photon that has probed an object is likely to strike the sensor~\cite{morris2015imaging}. This heralding has the advantage of removing some of the sensor and environmental parasitic noise, enabling the acquisition of images containing only a few photons~\cite{morris2015imaging}. A similar technique was used in the context of phase and amplitude imaging~\cite{Aidukas2018Ptycho}. The same technique is anticipated to be implemented in the context of LiDAR systems to generate quantum rangefinders~\cite{lanzagorta2011quantum}. In the context of a LiDAR system, the ability to overcome some of the technical noise can help such system to work with a fewer photons with the aim to see without being seen. Moreover, such schemes could be further improved by ad hoc implementations of the quantum illumination protocol proposed in the context of quantum information schemes~\cite{lloyd2008enhanced}.

\subsection*{Superresolution in quantum imaging.}
The quantum improvements in imaging are not only limited to noise decreasing. Resolution advantages can also be obtained by using quantum behaviour of light. Two main methods have been devised that harness quantum states to go beyond the diffraction limit.\\
The first method allows the standard quantum limit in resolution to be reached and goes beyond the diffraction limit by detecting quantum correlations between N photons; such a limit scales as $\frac{1}{\sqrt{N}}$ (Ref.~\cite{giovannetti2009sub}). It was shown that in this context of raster-scan imaging techniques it is possible to post-select a number of photons within classical light focused on an object to access the same kind of advantage~\cite{guerrieri2010sub,mouradian2011achieving}.\\
It is also possible to obtain a quantum enhancement in resolution with classical illumination when imaging fluorescent single photon emitters~\cite{schwartz2012improved}.
The light emitted by such objects exhibits photon anti-bunching, which can be harnessed to obtain a standard quantum limited resolution enhancement.
Experimental realizations have been performed to demonstrate resolution improvement using colloidal quantum dots~\cite{schwartz2013superresolution,israel2017quantum} and nitrogen-vacancy colour centres in diamond in the context of fluorescence confocal microscopy~\cite{monticone2014beating}. Interestingly, such techniques can be used in conjugation with classical techniques allowing super-resolution imaging to be achieved~\cite{classen2017superresolution}. This was demonstrated experimentally on biological samples stained with fluorescent quantum dots by combining the anti-bunching resolution advantage of quantum dots with the classical advantage of structured illumination~\cite{tenne2018super}. Finally it was shown that using photon counting strategies with classical illumination can lead to the resolution enhancement even for the observation of non-fluorescent objects~\cite{tsang2016quantum}.\\
In the context of full field imaging of non fluorescing objects (that is, without scanning), states exhibiting quantum correlated illumination can be used to gain a resolution improvement~\cite{giovannetti2009sub}. Recently, we have demonstrated a resolution enhancement in full-field imaging under of non fluorescing objects \cite{toninelli2019resolution} using a centroid measurement detection method for bi-photons~\cite{tsang2009quantum}.

\begin{figure}[H]
\centering
\includegraphics[width=\textwidth]{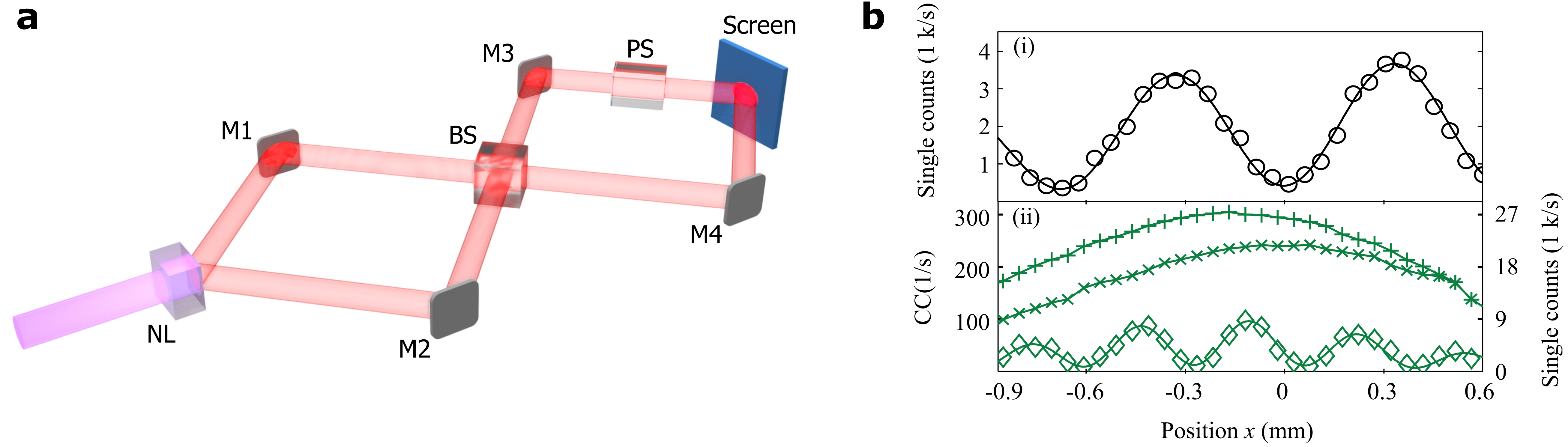}
\caption{\textbf{Principle of quantum lithography.} \textbf{a} Simplified version of a quantum lithography experimental scheme. A NOON state with N=2 is generated at the output of a beam-splitter (BS) through a HOM effect by recombining two photons generated by SPDC. The two photons are then in a superposition state of being both in the upper arm or in the lower arm of the interferometer situated between the BS and the screen. A phase shifter (PS) can be added in one of the paths. And one can displace two adjacent detectors in the plane of the screen to detect two photon coincidences and compute the centroid of such detected bi-photons. $M_{1,2,3,4}$ are mirrors, NL is a non-linear crystal.  \textbf{b} Experimental results evidencing the principle of quantum lithography. The black circles correspond to the single count interference pattern obtained with a classical coherent state. In the bottom panel, the two red curves correspond to single counts detected at the two detectors (right arrow). The green curve correspond to the coincidence count (CC)' centroid measurements (left arrow). One can observe that the period of the interference fringes on this later figure is approximately half that of the classical interference pattern. Panel b is reproduced from REF.~\cite{shin2011quantum}, APS.}
\label{fig:litho}
\end{figure}

The second method allowing Rayleigh's limit of diffraction to be surpassed is called quantum lithography. It uses the interference exhibited by NOON states to reach an improvement which scales as $\frac{1}{N}$ (Heisenberg scaling) in the size of projected interference fringes~\cite{boto2000quantum}. To access such an Heisenberg scaling it is required that the ensemble of photons `simulate' the behaviour of an indissociable photon that in an interferometer is in a superposition of being found in one arm or in the other. The equivalent situation with N photons is therefore for them to be in a superposition of being found altogether in one arm of the interferometer or to be all found in the other arm. Such a state is exactly a NOON state. In this situation, the super resolution phenomenon emerges in the N photon interference pattern so that, the pattern requiring a multi-photon absorption process is detected or printed on a material (see Fig. \ref{fig:litho}).\\
The difficulty in finding materials or detectors that are capable of performing multi-photon absorption has considerably limited the use of these two methods, and also the performance of the earliest realizations of quantum lithography~\cite{d2001two,chang2006implementation}. However it was suggested that to detect such interference patterns one could simply proceed to optical centroid measurements of the N detected photons of the NOON state~\cite{tsang2009quantum}. 
This method was later implemented~\cite{shin2011quantum} for 2 photon NOON states and for up to 4 photon NOON states~\cite{rozema2014scalable,matthews2014scalable}. Finally, it has been shown that multiphoton interference with independent single photon light sources can also lead to a similar super-resolved interference pattern~\cite{thiel2007quantum}, without using path entangled states. This is enabled by an effect equivalent to the extension of the Hanbury Brown–Twiss effect for intensity correlations with more than 2 photons~\cite{oppel2012superresolving}.

\section*{New imaging techniques}

In addition to improving the images obtained through conventional imaging techniques quantum effects have also allowed new methods of imaging to be developed. We report some the examples, including the ghost imaging. This technique arguably initiated the field of quantum imaging in that it was the first use of light containing spatial quantum correlations to illuminate the object to be imaged.

\subsection*{Ghost imaging}

Ghost imaging utilizes correlations between spatially separated light fields to record an image of an object using photons which have not interacted with the object~\cite{moreau2018ghost}. This technique uses quantum-entangled photons produced via SPDC within a non-linear medium and exploits quantum spatial correlations between the photon pairs that comprise the signal and idler output. The ghost imaging scheme was proposed originally in the late 1980's~\cite{klyshko_simple_1988} and then first realized in 1995 (Ref.~\cite{strekalov_observation_1995,pittman1995optical}). A typical ghost imaging setup comprises the spatial separation of the signal and idler photons, where the idler interacts with the object and is subsequently detected by a non-spatially resolving (bucket) detector; coincidence detection of a idler photons at the bucket detector and signal photons at the imaging detector is then performed to select the correlated photon pairs to build an image of the object. Of the detected signal and idler light fields neither beam alone contains the information required to reconstruct an image of the object. The bucket detector is spatially unresolved simply detecting all the idler photons passed by the object. Similarly the spatially resolving detector measures the position of all the signal photons incident upon it without any information about the object. However, the correlation between these datasets can be exploited to extract an image. Through integrating over a number of acquired events an image of the object may be recovered from the subset of the signal light field that is correlated with the idler light field detected by the bucket detector. Although initial realizations of ghost imaging entailed the use of a raster scanning technique~\cite{strekalov_observation_1995,pittman1995optical,pittman_two-photon_1996}, more recent ghost imaging schemes utilize a spatially resolved detector --- such as a time-gated ICCD camera triggered by a single photon avalanche diode --- to enable imaging across the full field of view without the inherent limitation in detector efficiency of $1/n$ for $n$ pixels for implementations based on scanning ~\cite{aspden_epr-based_2013,morris2015imaging,aspden_photon-sparse_2015}.\\
After the earliest realizations of ghost imaging~\cite{strekalov_observation_1995,pittman1995optical}, it was unclear to the community if entanglement was actually needed to perform ghost imaging. It was, however, later demonstrated that it is possible to perform ghost imaging by using a classical source which is a laser beam, deflected by a variable amount and then passed through a beam splitter~\cite{bennink2002two} or thermal light split in two beams on a beam splitter~\cite{gatti2004ghost}. Regarding the potential advantages of quantum versus classical ghost imaging, it is now recognized that these methods produce images of a similar resolution~\cite{gatti2008quantum,moreau_resolution_2018}. The main advantage of quantum light is found at low light levels, where it exhibits greater visibility and a greater signal to noise ratio~\cite{gatti2008quantum,brida2011systematic}. Nevertheless, classical ghost imaging techniques have inspired new type of imaging based on the use of classical correlations~\cite{baleine2006correlated,pepe2017diffraction,altmann2018quantum}. For an overview of the comparison between classical and quantum ghost imaging, see \cite{gatti2008quantum,shapiro2012physics,genovese2016real}.\\

A degenerate quantum ghost imaging system in which the signal and idler photons are of the same wavelength has applications for the imaging of samples under low light conditions~\cite{morris2015imaging}, for example, in compressed sensing and object tracking~\cite{magana-loaiza_compressive_2013}. However, it is possible to design a non-degenerate ghost imaging system in which the signal and idler photons are of different wavelength. The use of such non-degenerate down-conversion source enables the imaging of objects in wavebands in which spatially resolved detectors are impractical, expensive or ineffective in terms of resolution and, therefore, cannot be used in imaging applications. To perform ghost imaging in such wavebands only a bucket detector is required while the imaging detector operates in a waveband where spatially resolved detectors are relatively efficient and inexpensive.

\begin{figure}[H]
\centering
\includegraphics[width=\textwidth]{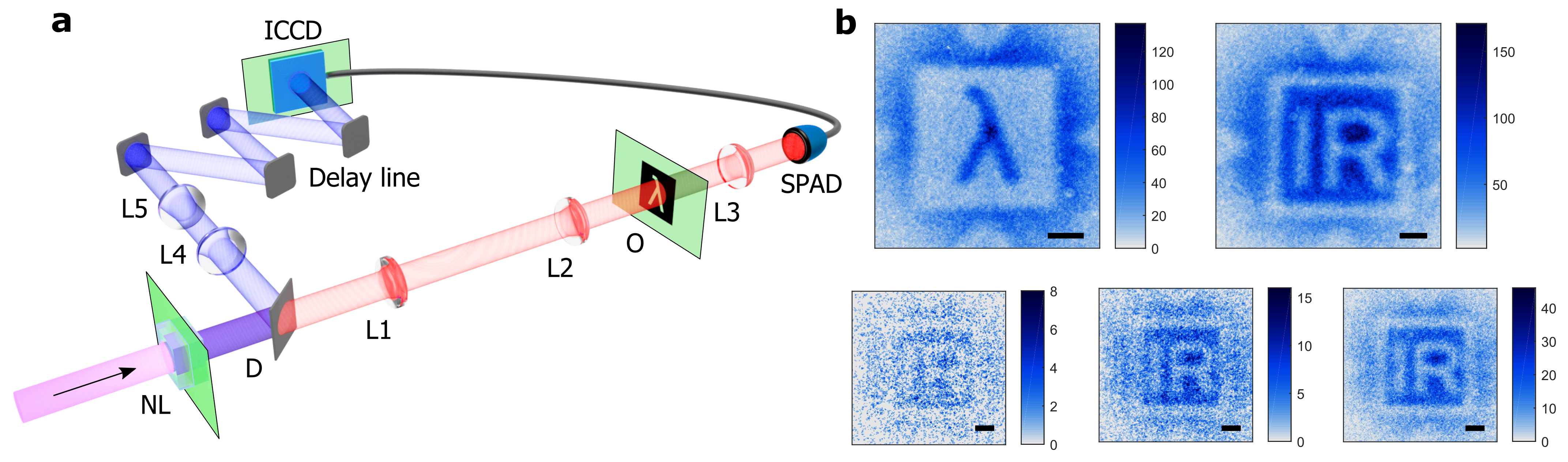}
\caption{\textbf{Non-degenerate quantum ghost imaging.} \textbf{a} Model of the experimental setup for non-degenerate quantum ghost imaging. The non-degenerate signal and idler beams generated in a barium borate crystal cut for Type I phase matching (NL) are separated at the dichroic mirror (D). The infrared idler photon interacts with the object and is detected by a non-spatially resolving SPAD. The corresponding signal photons detected by an intensified CCD (ICCD) camera. The ICCD camera is triggered by the arrival of a photon at the single-photon avalanche diod (SPAD) detector. $L_{1,2,3,4,5}$ are the lenses and O is the imaging object. \textbf{b} Images of objects created by gold pattered onto a silicon wafer by electron beam lithography. Contrary to silicon, gold is non-transmissive at the idler wavelength of 1550 nm. Both materials are non-transmissive at the signal wavelength of 460 nm. Ghost images of the object in the IR are acquired. The three images in the bottom row are acquired with increasing number of photons. Scale bars, 50$\mu m$, the colour bar represents the number of counts. Panel b is reproduced from REF.~\cite{aspden2015photon}, OSA.}
\label{fig:transwavelengthGhostImaging}
\end{figure}

Non-degenerate ghost imaging has been carried out in 2015 (Ref~\cite{aspden2015photon}) using an experimental setup as modelled in {Fig. \ref{fig:transwavelengthGhostImaging}a}. In this setup the signal and idler photons at 460 nm and 1550 nm respectively are separated at the dichroic mirror D; an object with features observable in the infrared is probed by the idler photons in the ghost imaging arm and detected by the non-spatially resolved SPAD, while the signal photons are detected using the spatially resolved ICCD camera. Non-degenerate ghost imaging allows to visualize an object in the infrared domain, using photons of a visible wavelength to reconstruct the image (see Fig. \ref{fig:transwavelengthGhostImaging}b).

In the aforementioned work~\cite{aspden2015photon} the objects were gold patterned onto silicon wafer using electron beam lithography, of which the silicon is transmissive for the infrared photons while the gold is not (\textbf{Figure \ref{fig:transwavelengthGhostImaging}b}). Non-degenerate ghost imaging using idler wavelengths outwith the current spectral range of spatially resolved detectors potentially would allow the imaging of objects and their internal structure using far-infrared and terahertz wavelengths. These wavelengths have applications in biological, industrial, and security imaging applications in which raster scanning techniques using the aforementioned non-spatially resolved detectors are too slow in the construction of an image~\cite{jansen_terahertz_2010}. As for the resolution of non-degenerate ghost imaging, theoretical works indicate that the resolution does not lead to improvement over the classical methods~\cite{rubin_resolution_2008,chan_two-color_2009}; also the resolution will still be limited by the point spread function of the imaging system and may be degraded by reducing the strength of correlation between the photon pairs~\cite{moreau_resolution_2018}.\\
Finally, quantum correlated sources have been used to perform ghost acquisitions in other domains. For example, temporal ghost imaging has been performed using optical correlations~\cite{denis2017temporal}. However, it also found its usage outside of conventional optics with recent demonstrations performed in the X-ray domain~\cite{schori2018ghost} and using beams of correlated atoms~\cite{khakimov2016ghost}.

\subsection*{Imaging with undetected photons}

\begin{figure}[H]
\centering
\includegraphics[width=\textwidth]{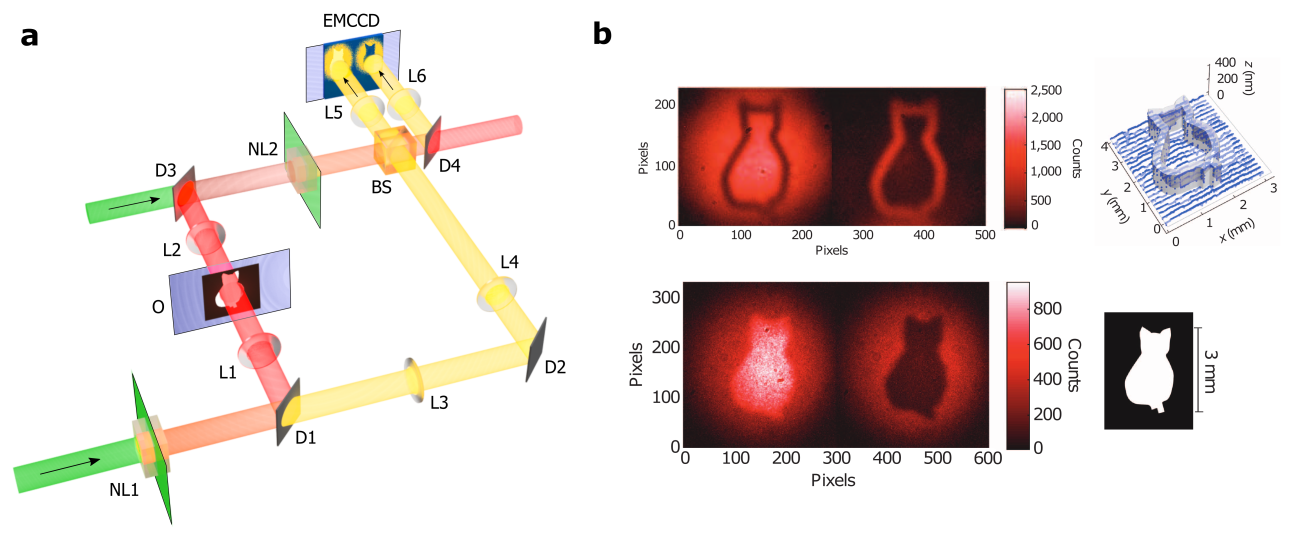}
\caption{\textbf{Quantum imaging with undetected photons.} \textbf{a} Model of the experimental setup used to perform imaging without detecting the photons that interact with the object. A non-linear interferometer is built by seeding the idler produced in the first non-linear crystal NL1 into a second one NL2 and recombining the signal waves emitted by the two crystals on a beam splitter (BS). Due to quantum interference between the two photons wave-functions emitted by the crystals, an interference figure appears within the signal intensities outputting the BS. If one adds an object between NL1 and NL2 in the idler beam, one affects the whole wave function and as a consequence, the interference of the signal waves after the BS. One can in such conditions acquire an image of the object without having to detect the photons that have interacted with it. $L1,2,3,4,5,6$ are lenses, $D1,2,3,4$ are the dichroic mirrors, O is the imaging object and EMCCD is the electron-multiplying CCD camera. \textbf{b} Images of an object obtained by detection of the interference of signal photons that have not interacted with it. The idler photons that have interacted with the object remain undetected. The top row represents the imaging of a phase object, the blue dots on the right plot correspond to a scan of the object etch depth, the bottom row represents the imaging of a transmissive object. The colour bar represents the number of counts. Panel b is reproduced from Ref.~\cite{lemos2014quantum}, Springer Nature Limited.}
\label{fig:undetectedPhotonsImaging}
\end{figure}

Further to the aforementioned method of ghost imaging --- in which the detection of the idler photon is used to herald the arrival of its twin on an imaging detector and to select a subset of the signal photons in order to produce an image --- it has been shown ~\cite{lemos2014quantum,lahiri_theory_2015} that it is not necessary to detect the idler photon at all to form an image using the signal photons. The physical basis of this technique was first introduced and demonstrated in the context of mono-mode quantum interference~\cite{zou_induced_1991,wang_induced_1991}. In these works it was shown that a non-linear quantum interferometer can be built by using two non-linear crystals, and that by feeding the idler wave emerging from the first non-linear crystal into the second crystal, one can induce coherence between the two signal beams emitted by the crystal without inducing emission (that is,  with amplification of the idler wave).\\
In these experiments a laser beam is split in two beams that pump coherently a pair of non-linear crystals which are phase-matched for non-degenerate down-conversion. The signal and idler waves emitted through SPDC within the first crystal (NL$_1$) are separated and the idler wave interacts with an object ({ Fig. \ref{fig:undetectedPhotonsImaging}}). 

After interaction with the object, the idler is then directed into the second crystal (NL$_2$), and co-axially aligned with the spontaneous idler emanating from it. The signal wave at the output of NL2 is then separated from the idler waves that originate from either NL1 and NL2, and the signal waves from the pair of SPDC events, neither of which have interacted with the object, are combined at a beam-splitter. In such conditions the signal waves, even though they have not directly interacted with the object, will undergo interference dependant on the phase and transmission of the object as probed by the idler originating from NL1.\\
To understand this phenomenon one can first remark that the superposition of the two idler waves generated in the two crystals makes it impossible to distinguish in which crystal an idler photon was created. This has the potential to induce coherence between the bi-photon states emitted by the two crystals. It can happen even without induced emission in the second crystal that could be the consequence of an idler field at its input~\cite{wang_induced_1991,ou_coherence_1990}. When the two signal waves are combined on a beam-splitter, there is no way to distinguish from which crystal either of the signal and idler photons comes from and interference patterns that appear are detectable directly on the intensities measured on the signal side. When an absorptive object is a added between the two crystals on the idler beam path, some distinguishability is introduced and the interference visibility of the signal is reduced as the coherence is partially lost.\\
Using this method in the context of imaging, it was shown~\cite{lemos2014quantum} that both intensity and phase objects can be imaged without having to detect the idler waves that have interacted with the object (note, however,that the object is still illuminated). They were able to show imaging of a phase object which is opaque to the wavelength of the detected signal photons (820 nm) where the object was probed by an idler wave at which it is transparent (1515 nm), {Fig. \ref{fig:undetectedPhotonsImaging}}. This capacity to image with undetected photons possesses an advantage over non-degenerate ghost imaging techniques due to there being no requirement for a detector of any sort at the probe wavelength thereby overcoming constraints with regards to detector availability and detection efficiency within certain wavebands. A similar method has been applied to spectroscopy where the spectrum of a sample in the infra-red is obtained by detecting visible photons only~\cite{kalashnikov2016infrared}. The method may also find application to quantify correlations between two beams without having to detect each of the two beams~\cite{hochrainer2017quantifying}. However, the fact that such a technique involves interferometry makes it more complex to implement and more susceptible to mechanical and thermal noise.\\

\subsection*{Interaction-free measurement} 

Previous methods have detailed the cases where photons interacted with the object of interest and using either the time or position of the photon detection and of its entangled partner to obtain information about the object that has been imaged. However, it is also possible to acquire information regarding an object without a photon having interacted with it by performing interaction-free measurements.\\
The concept of interaction-free measurement was introduced in 1993 (Ref.~\cite{elitzur1993quantum, vaidman_l._realization_1994}).
In this experiment, a Mach-Zehnder interferometer is set up with the two detectors D1 and D2 that are positioned for constructive and destructive interference respectively. Any single photons injected into the system should therefore be detected at D1 and none at detector D2. A perfect absorbing object placed into one of the arms would be transmitting a single photon through the system. Should a single photon be detected at detector D2 then there must have been an object in one arm of the interferometer. However, as the single photon was detected it may not have interacted with the object and passed through the unobstructed arm. It is to note that despite its name, interaction-free measurement is not truly interaction-free as there will be a coupling in any quantum mechanical or wave-like description of the system~\cite{kwiat_experimental_1998,vaidman_are_2001,geszti_interaction-free_1998}.\\

In their striking thought experiment, often referred to as the Elitzur–Vaidman bomb tester, Elitzur and Vaidman used bomb which would explode should it detect a photon as the object into one of the arms~\cite{elitzur1993quantum, vaidman_l._realization_1994}. Should the Mach-Zehnder interferometer be balanced then success rate of finding bomb without exploding it would be $\eta = 1/3$. This rate is also referred to as the `interaction free efficiency'. However, for an asymptotically fully unbalanced interferometer, this efficiency can tend toward $\eta = 1/2$. Although for a standard interferometer this is the efficiency limit, it is possible to use a different type of interferometer to implement an optical analogue of the quantum Zeno effect \cite{misra_zenos_1977}, that allows to create a lossless system with respect to the `bombs', with  an efficiency tending to $\eta = 1 $. This was proposed in 1995 (Ref.~\cite{kwiat_p_interaction-free_1995}) and it is based on the principle of performing successive weak measurements of the presence of the object.\\

The Elitzur-Vaidman interaction-free detection scheme was experimentally demonstrated by using a SPDC source of photon pairs, one of which enters an unbalanced Michelson interferometer while the other is detected to herald the presence of the down-converted photon pairs leading to efficiencies close to $\eta = 1/2$ (Ref.~\cite{kwiat_p_interaction-free_1995}). The quantum Zeno effect version was realized within a Fabry-Perot interferometer with a detection efficiency of $85\%$ (Ref.~\cite{tsegaye_efficient_1998}) and through a looped polarisation sensitive interferometer~\cite{kwiat_high-efficiency_1999,kwiat_experimental_1998} with an efficiency of $73\%$.

Interaction-free scanned imaging, has been performed~\cite{white1998interaction} by scanning an object at a focus point of the beam within a polarization sensitive Mach-Zehnder interferometer for the purposes of determining the dimensions of objects placed in one of the arms. In this experiment a number of objects were scanned across the beam and the dimensions of said objects, of the order $10-100~\mu m$, were accurately assessed. Recently, the interaction-free imaging of a structured object has been demonstrated using a quantum ghost-imaging setup~\cite{zhang2019interaction}.

\section*{Outlook}

As we have seen, quantum imaging has allowed to test quantum mechanic phenomena and also enabled the development of new imaging protocols. Quantum imaging has also contributed to the emergence of new `quantum inspired' imaging protocols like classical ghost imaging~\cite{gatti2008quantum} or single pixel camera implementations~\cite{altmann2018quantum}. The emergence of new, more efficient photon-counting cameras --- such as superconducting nano-wires, single photon detector arrays, very low noise back-side-illumination CMOS technologies, and new sources like quantum dots and exotic non-linear sources, that are currently under development --- gives confidence regarding the future of quantum imaging schemes that should soon reach the performance levels required to deliver practical implementations. An example would be the development of sources of pairs of photons with extremely different wavelengths thereby overcoming the lack of high--fidelity detectors at exotic wavelengths. Such sources could be used advantageously for either imaging without detection scheme or a non-degenerate ghost imaging scheme where an interferometric scheme would be difficult to implement. The developments that are currently ongoing have the potential to bring efficient imaging techniques and sensors to new domains of optics. Further to the potential applications of quantum imaging schemes, it has been demonstrated that imaging schemes at conventional wavelengths can be improved by using quantum sources, either by enhancing the resolution or decreasing the noise of images~\cite{toninelli2019resolution,brida2010experimental}. Future developments in quantum imaging could come through new approaches such as quantum image processing~\cite{yan2017quantum} or through exploiting quantum correlations via different methods such as correlation plenoptic imaging~\cite{di2018correlation}. Moreover, early demonstrations of principle of quantum imaging techniques are now more and more transforming into real world practical implementations from which new imaging technologies may emerge. Finally, the high dimensionality accessible in the space domain through imaging together with the increase of the efficiency of quantum imaging protocols should in the future enable quantum information protocols based on imaging that will allow the design of `inherently efficient' information protocols in high dimensions. At a stage where worldwide research funding bodies are pushing for the development of real world quantum technologies through the so-called second quantum revolution~\cite{schleich2016quantum,barnett2017journeys,mohseni2017commercialize}, we believe that quantum imaging will have an important role to play in this movement.

\bibliographystyle{ieeetr}

\section*{Acknowledgements}
This work was funded by the UK EPSRC (QuantIC EP/M01326X/1) and the ERC (TWISTS, 340507, Grant no. 192382). P.-A. M. acknowledges the support from the European Union's Horizon 2020 research and innovation programme under the Marie Sklodowska-Curie fellowship grant agreement No 706410, of the Leverhulme Trust through the Research Project Grant ECF-2018-634 and of the Lord Kelvin / Adam Smith Leadership Fellowship scheme. E.T. acknowledges the financial support from he EPSRC Centre for Doctoral Training Intelligent Sensing and Measurement (EP/L016753/1). T.G. acknowledges the financial support from the EPSRC (EP/N509668/1) and the Professor Jim Gatheral quantum technology studentship.
\section*{Author contributions}
P.-A.M. and M.J.P. made substantial contributions to discussions of the content.
P.-A.M., T.G. and M.J.P. researched data for the article.
P.-A.M., E.T., T.G. and M.J.P. wrote the article and reviewed and/or edited the manuscript before submission.

\end{document}